\documentclass[sigconf]{acmart}

\AtBeginDocument{%
  }

\usepackage{algpseudocode}
\usepackage{algorithm}
\usepackage{caption}
\usepackage{amsmath}
\usepackage{amsfonts}
\usepackage{subcaption}
\usepackage{mdframed}
\usepackage{listings}
\usepackage{makecell}
\usepackage{xcolor}
\usepackage[most]{tcolorbox}
\usepackage{fontawesome}
\usepackage{color}
\usepackage{xcolor}
\usepackage[normalem]{ulem}
\usepackage{multirow}
\usepackage{wrapfig}
\usepackage{relsize}
\usepackage{url}
\usepackage{xspace}
\definecolor{gray}{rgb}{0.4,0.4,0.4}
\definecolor{darkblue}{rgb}{0.0,0.0,0.6}
\definecolor{cyan}{rgb}{0.0,0.6,0.6}

\lstdefinelanguage{XML}
{
  morestring=[b]",
  morestring=[s]{>}{<},
  morecomment=[s]{<?}{?>},
  stringstyle=\color{black},
  identifierstyle=\color{darkblue},
  keywordstyle=\color{cyan},
  morekeywords={xmlns,version,type}
}

\title{On the Freshness of Pinned Dependencies in Maven}

\author{Vasudev Vikram}
\affiliation{%
  \institution{Carnegie Mellon University}
  \city{Pittsburgh}
  \country{USA}
}
\email{vasumv@cmu.edu}

\author{Yuvraj Agarwal}
\affiliation{%
  \institution{Carnegie Mellon University}
  \city{Pittsburgh}
  \country{USA}
}
\email{yuvraja@cs.cmu.edu}

\author{Rohan Padhye}
\affiliation{%
  \institution{Carnegie Mellon University}
  \city{Pittsburgh}
  \country{USA}
}
\email{rohanpadhye@cmu.edu}

\newcommand\library[2]{{#1}@{#2}}
\newcommand\version[2]{{#1}^{#2}}
\newcommand\upgrade[2]{$\langle{#1}, {#2}\rangle$}
\newcommand\pin[3]{$\langle${#1}, {#2}, {#3}$\rangle$}
\newcommand{\change}[1]{#1} 
\newcommand{\delete}[1]{} 
\newcommand{\tool}{Pin-Freshener\xspace}

\definecolor{boxcolor}{RGB}{238, 223, 204} %
\DeclareRobustCommand{\mybox}[2][gray!20]{%
\begin{tcolorbox}[   
        breakable,
        left=0pt,
        right=0pt,
        top=0pt,
        bottom=0pt,
        colback=#1,
        colframe=black,
        width=\dimexpr\columnwidth\relax, 
        enlarge left by=0mm,
        boxsep=5pt,
        outer arc=4pt,
        boxrule=.5mm
        ]
        #2
\end{tcolorbox}
}

\setcopyright{none} 
\renewcommand\footnotetextcopyrightpermission[1]{}

\begin{document}
\sloppy
\begin{abstract}
Library dependencies in software ecosystems play a crucial role in the development of software. As newer releases of these libraries are published, developers may opt to \textit{pin} their dependencies to a particular version. While pinning may have benefits in ensuring reproducible builds and avoiding breaking changes, it bears larger risks in using outdated dependencies that may contain bugs and security vulnerabilities. To understand the frequency and consequences of dependency pinning, we first define the concepts of \textit{stale} and \textit{fresh} pins, which are distinguished based on how outdated the dependency is relative to the release date of the project. We conduct an empirical study to show that over 60\% of consumers of popular Maven libraries contain stale pins to their dependencies, with some outdated versions over a year old. These pinned versions often miss out on security fixes; we find that 10\% of all dependency upgrades in our dataset to the latest minor or patch version would reduce security vulnerabilities. 

We prototype an approach called \tool{} that can encourage developers to freshen their pins by leveraging the insight that \textit{crowdsourced} tests of peer projects can provide additional signal for the safety of an upgrade. Running \tool{} on dependency upgrades shows that just 1--5 additional test suites can provide 35--100\% more coverage of a dependency, compared to that of a single consumer test suite. Our evaluation on real-world pins to the top 500 popular libraries in Maven shows that \tool{} can provide an additional signal of at least 5 passing crowdsourced test suites to over 3,000 consumers to safely perform an upgrade that reduces security vulnerabilities. \tool{} can provide practical confidence to developers by offering additional signal beyond their own test suites, representing an improvement over current practices.
\end{abstract}

\maketitle
\section{Introduction}
\begin{figure}[t]
    \centering
    \includegraphics[width=\columnwidth]{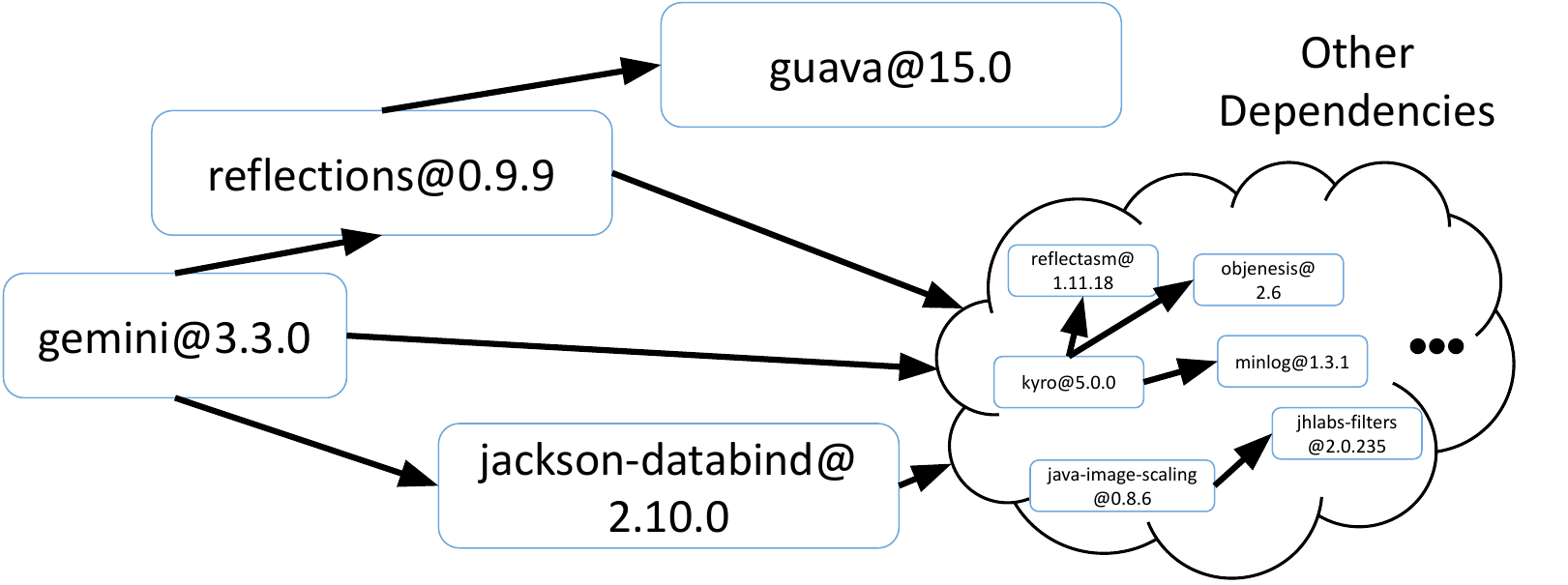}
    \caption{Example dependency tree of the Maven library \texttt{gemini@3.3.0}. A directed arrow denotes a dependency. Each node consists of a library name and version. \texttt{gemini@3.3.0} contains a direct dependency to \texttt{jackson-databind@2.10.0} and an indirect dependency to \texttt{guava@15.0}.}
    \label{fig:dependency-tree}
\end{figure}

Modern software heavily relies on third-party libraries. Usage of these libraries can reduce software development time and cost by reusing existing functionality of software~\cite{de2008empirical, cataldo2009software}. This process has been integrated into many software ecosystems---such as Apache Maven for Java, NPM for JavaScript, and PIP for Python---for which building and installing library dependencies is a natural step for software developers. The Maven Central Repository demonstrates the popularity of this practice for Java applications, with an index containing millions of Java packages~\cite{mavencentral22}. An example of the dependency network of the Maven \texttt{gemini} library is shown in Figure~\ref{fig:dependency-tree}, showing many dependencies than can span multiple edges. 

While the dependence on third-party libraries assists the development of new software applications, managing these dependencies can be challenging. New releases of dependencies are constantly published to the ecosystem and developers must decide whether to upgrade them to a newer version. However, software bugs or unexpected behavior---referred to as breaking changes---can be introduced in these new versions~\cite{raemaekers2014semantic, xavier2017historical, ochoa2022breaking}. Third-party library maintainers sometimes even \textit{knowingly} deploy breaking changes due to the build up of technical debt and pressure to release new functionality~\cite{bogart2016break}. 

Thus, upgrading a dependency can always be risky for consumers of these libraries. They may be wary of the possibility that their project might break or even that new security vulnerabilities are introduced~\cite{ponta2020detection}. Developers are often faced with the choice of \textit{floating} their dependencies to a range of versions (e.g. \texttt{2.1+}), or \textit{pinning} their dependencies to a single specific version (e.g. \texttt{2.1}). Software ecosystems also adopt different conventions for specifying dependency versions; the default convention in NPM is to specify a floating version, whereas the default in Apache Maven is to specify a fixed pinned version. In the case of Maven, this is largely due to the packages not necessarily following the semantic versioning practices when new versions are released~\cite{raemaekers2014semantic}.


Thus, pinning has certain benefits such as providing reproducible builds~\cite{mukherjee2021fixing} and avoiding breaking changes, but it can also bear a significant cost! New library versions often include new features, performance improvements, and crucial security patches. The high-profile 2017 Equifax data breach, in which a vulnerability in the open source Apache Struts library was exploited for leaking sensitive data of over 140 million consumers, demonstrates this drawback of pinning~\cite{CVE-2017-5638}. A patch for Apache Struts was available, but was not adopted by Equifax for over \textit{two months}. Nowadays, tools like \emph{Dependabot} and others~\cite{Dependabot, alfadel2021use, he2023automating, mohayeji2023investigating} help warn developers about known security vulnerabilities in outdated dependencies, though this approach is reactive rather than proactive.

So, we ask: is dependency pinning actually worth it? \change{We first distinguish between two types of pins: \textit{stale} pins and \textit{fresh} pins. A \textit{stale} pin captures an explicit instance in which a project, at the time of release, was pinned to an outdated dependency even though a newer one was available. On the other hand, a \textit{fresh} pin is one in which a project is pinned to the latest version of the dependency available.} Using these definitions, we conduct an empirical study on the Maven ecosystem to understand the how common the practice is and its broader security implications. We use the Open Source Insights dataset~\cite{DepsDev}, published by Google, containing data about dependencies, consumers, and security vulnerabilities for over 569,000 Maven packages. We construct a dataset from a targeted sample of the most popular Maven libraries from the Open Source Insights dataset and find that \textit{over 60\%} of the consumers of these libraries contain stale pins to outdated versions. 

Given that stale dependency pinning is a fairly common practice in Maven, we next explore the security risks it poses. In our historical analysis on pinned dependencies, we find that 10\% of library upgrades could have fixed security vulnerabilities had they adopted \textit{fresh pinning} for their dependencies when publishing their library. In contrast, only 3\% of the upgrades would have introduced new vulnerabilities. This corresponds to over 22,000 consumers in our dataset that potentially could have fixed vulnerabilities (a majority of which having high or critical severity levels) had they performed upgrades to fresh pins. Hence, we believe that developers should follow a more proactive pinning strategy since they are far likelier to fix vulnerabilities by upgrading their outdated dependencies. We acknowledge that pinning has benefits such as maintaining reproducibility of builds and avoiding unexpected breakage; however, the prevalence of stale pins and their security implications suggests developers should adopt a fresh pinning strategy. 

While the overall security benefit of upgrading stale pins is clear, we must still consider the aspect of evaluating whether performing a specific upgrade is safe. In Maven, breaking changes even across minor version updates are fairly common~\cite{raemaekers2014semantic}. So, how can we provide more confidence for consumers upgrade their stale pins to fresh pins? Our key insight is that the test suites of other consumers in the ecosystem can help provide additional signal about the upgrade and more confidence to the developer. To this end, we develop a prototype of \emph{\tool{}}, an approach that \textit{crowdsources} test suites of peer consumers of the dependency to evaluate the safety of an upgrade to a fresh pin. We specifically leverage the existence of \textit{test-JARs} in the Maven ecosystem, which contain projects' compiled tests, in order to streamline the execution of consumer test suites. By executing these additional test suites against both the pinned version and upgraded version, we can characterize the impact of the upgrade on multiple projects. \tool{} reports a \textit{confidence score} of a particular upgrade determined by the number of consumer test suites that are able to successfully run when using the upgraded dependency version.

In an experiment of executing \tool{} on our dataset of these upgrades, we first find that crowdsourcing just five consumer test suites is able to provide an average improvement of almost 100\% in terms of test coverage of a dependency over that of a single consumer. \tool{} is able to provide an additional signal of at least five passing crowdsourced test suites to over \textit{3,000} consumers (15\%) performing upgrades that would fix security vulnerabilities. Although this signal does not provide \textit{guarantees} of upgrade safety, it can provide developers with additional practical confidence (similar to code coverage metrics) beyond what they would get from relying solely on their own tests--the current standard practice for judging whether to perform an upgrade.

In summary, we ask the following research questions:
\begin{description}
\setlength{\itemsep}{0pt}
\setlength{\parskip}{2pt}
    \item[\textbf{RQ1:}]To what extent are libraries in the Maven ecosystem pinning to stale dependencies?

    \item[\textbf{RQ2:}]What is the security impact of pinning to stale dependencies?

    \item[\textbf{RQ3:}]How much can crowdsourced test suites improve coverage of the pinned dependency?

    \item[\textbf{RQ4:}]Can crowdsourced test suites help provide additional signal for vulnerability-fixing upgrades from stale pins to fresh pins?
\end{description}

\noindent Our contributions are as follows:
\begin{enumerate}
\item We introduce the concepts of a \textit{stale} and \textit{fresh} pins, which depend on whether a project is pinned to an outdated dependency. Using this definition, we conduct an empirical study on the Apache Maven ecosystem using the Open Source Insights dataset to determine the frequency and security impact of dependency pinning relating to the top 500 most-popular libraries. 
\item We \change{design and implement a prototype of \emph{\tool{}}, an approach that crowdsources consumer test suites to better characterize the safety of an upgrade across the network and provide additional signal to developers when freshening pinned dependencies.}
\item We run \tool{} at a large scale on vulnerability-fixing upgrades in Maven libraries and find that \tool{} is able to provide additional signal of at least 5 passing crowdsourced test suites to over 3,000 consumers that can perform an upgrade that reduces security vulnerabilities from a stale pin to a fresh pin.
\end{enumerate}

\section{Background and Terminology}
\label{sec:background}

\subsection{Dependencies in Software Ecosystems}
\label{sec:background:ecosystems}

A software ecosystem is a collection of software libraries, each denoted by a name and a version number. We denote a library as $\library{L}{\version{V}{}}$, where $L$ refers to the library name and $\version{V}{}$ refers to version. We define $\mathbb{L}$ as the set of all libraries in a particular software ecosystem, such as Maven for Java. 

A library $\library{L}{\version{V}{}}$ may contain a \textit{direct dependency} to another library $\library{L^\prime}{\version{V}{\prime}}$, usually specified in a configuration file for the build system. Throughout this paper, we refer to a dependency as the specific package as pulled by the build system after dependency resolution. The dependency resolution process will resolve any wildcard versions or ranges specified in the configuration file and fetch one single version of the dependency. We refer to $\library{L^\prime}{\version{V}{\prime}}$ as a \textit{direct dependency} and $\library{L}{\version{V}{}}$ as a \textit{direct consumer}. A shorthand notation for describing this direct dependency relation is $\library{L}{\version{V}{}} \rightarrow \library{L^\prime}{\version{V}{\prime}}$. An example of a direct dependency relation can be seen in Figure~\ref{fig:dependency-tree} between \library{\texttt{gemini}}{\texttt{3.3.0}} and \library{\texttt{jackson-databind}}{\texttt{2.10.0}}. We define the entire dependency graph $\mathbb{G}$ as the set of all direct dependency relations (edges), and naturally define the functions \textit{directDeps} and \textit{directConsumers} to identify a direct dependency on $D$ or a direct consumer $C$ respectively as follows:
\begin{align*}
    \textit{directDeps}(\library{L}{V}) &= \{\library{D}{\version{V}{\prime}} \in \mathbb{L} \mid \\
    &\qquad (\library{L}{\version{V}{}} \rightarrow \library{D}{\version{V}{\prime}}) \in \mathbb{G}\}\\
    \textit{directConsumers}(\library{L}{\version{V}{}}) &= \{\library{C}{\version{V}{\prime\prime}} \in \mathbb{L} \mid \\
    &\qquad (\library{C}{\version{V}{\prime\prime}} \rightarrow \library{L}{\version{V}{}}) \in \mathbb{G}\}
\end{align*}
A library dependency can also span multiple dependency edges, such as between \library{\texttt{gemini}}{\texttt{3.3.0}} and \library{\texttt{guava}}{\texttt{15.0}} in Figure~\ref{fig:dependency-tree}. To account for these dependency relations, we define the function \textit{allDeps} on $\library{L}{V}$ to return the transitive closure of \textit{directDeps} applied to $\library{L}{V}$. We similarly define \textit{allConsumers} as the transitive closure of \textit{directConsumers}. These functions return the set of all dependencies and consumers of $\library{L}{V}$, respectively, regardless of the number of edges. We additionally introduce the functions \textit{indirectDeps} and \textit{indirectConsumers} to return the sets of dependencies and consumers that are not direct. 

A library has the option of \textit{upgrading} a dependency from one version to a newer one. Continuing our example from Figure~\ref{fig:dependency-tree}, the library \library{\texttt{gemini}}{\texttt{3.3.0}} could upgrade \texttt{jackson-databind} from version \texttt{2.10.0} to \texttt{2.11.0}. We denote an \emph{upgrade} as the pair \upgrade{\library{D}{\version{V}{\alpha}}}{ \library{D}{\version{V}{\beta}}}, where $\version{V}{\alpha}$ and $\version{V}{\beta}$ are the older and newer versions of $D$, respectively.

\subsection{Semantic Versioning}
\label{sec:background:semver}

When performing a dependency upgrade, it's crucial for consumers to understand the types of changes being introduced in a new dependency version and whether it is backwards compatible. One practice used in many software ecosystems is \textit{semantic versioning}~\cite{semver}, which defines a set of rules for assigning version numbers to new releases of libraries. When using semantic versioning, a version $\version{V}{}$ is structured into the format \texttt{major.minor.patch[-tag]}. For example, the dependency \texttt{jackson-databind} in Figure~\ref{fig:dependency-tree} has version \texttt{2.10.0}, where 2 is the major version, 10 is the minor version, and 0 is the patch version. For notational purposes, we define the functions \textit{major}, \textit{minor}, and \textit{patch} to return the corresponding version numbers of a particular version $\version{V}{}$. This separation of version numbers also defines a total ordering between versions that compares major, minor, and patch versions numerically from left to right. We use this comparison logic throughout the paper when ordering versions (e.g. $\version{V}{\beta} > \version{V}{\alpha}$).   

Semantic versioning is used to characterize the types of version upgrades in terms of backwards compatibility. Generally, version upgrades that include backwards \textit{incompatible} changes increment the major version, whereas upgrades that do not break existing functionality are limited to minor or patch version increments. This allows library developers to notify consumers about the specific versions that introduce potential breaking changes, and consumers can choose which versions to adopt through a set of dependency constraints. Throughout this paper, we refer to minor and patch version upgrades as \textit{semver-compatible}, as they should have the assurance of being backwards compatible. 

Semantic versioning encourages consumers to perform semver-compatible upgrades on their dependencies since there should be no risk of introducing breaking changes. This can be as simple as specifying a version range for a dependency that freezes the major version, such as \texttt{[1.0.0, 2.0.0)}. However, semantic versioning is only a policy and is unenforceable throughout a software community; oftentimes new minor and patch versions may not respect the policy, resulting in unexpected breaking changes and upset consumers~\cite{decan19,wittern2016look}. These upgrades can even introduce accidental bugs or new security vulnerabilities, which may convince consumers to avoid semver-compatible upgrades entirely and decide to \textit{pin} their dependencies to a single version. \change{Package ecosystems also adopt different default approaches to versioning: NPM favors floating dependencies that follow semantic versioning conventions, automatically accepting compatible updates. Maven, on the other hand, defaults to pinning dependencies to specific versions, reflecting its ecosystem's less consistent adherence to semantic versioning principles. This is likely due to many libraries in Maven historically publishing breaking changes that disobey semantic versioning~\cite{raemaekers2014semantic}.}

\subsection{Dependency Pinning}

\label{sec:background:pin}
\begin{figure}[t]
    \centering
    \includegraphics[width=\columnwidth]{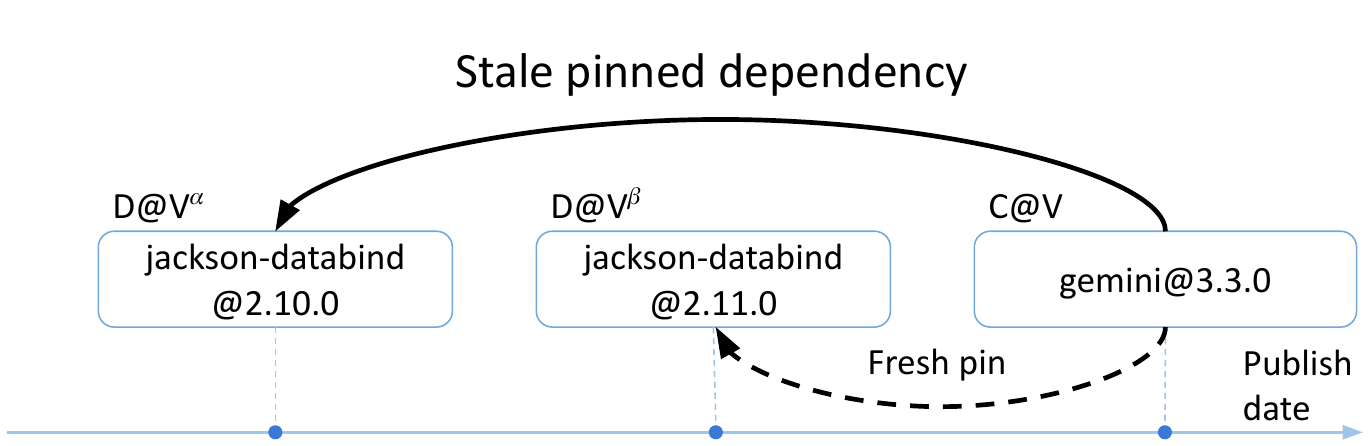}
    \caption{Example of a direct stale and fresh pin. Since $\library{\texttt{gemini}}{\texttt{3.3.0}}$ is a direct consumer of $\library{\texttt{jackson-databind}}{\texttt{2.10.0}}$ and the latest version of the library \texttt{2.11.0} was published before the consumer, this is a stale pin. If $\library{\texttt{gemini}}{\texttt{3.3.0}}$ was pinned to $\library{\texttt{jackson-databind}}{\texttt{2.11.0}}$, this would be a \textit{fresh} pin, as it is the latest version available at the time.}
    \label{fig:pinned-dependency}
\end{figure}



The practice of specifying a single version of a dependency rather than a range is referred to as \textit{dependency pinning}. Pinning to a specific version can provide benefits in ensuring reproducible builds and avoiding accidental breakage due to new breaking changes in an upgrade. However, the developer is then responsible for updating their pinned dependencies to the latest versions before releasing their own software. The failure to do so results in a \textit{stale} pin, which is an explicit instance in which software is pinned to an outdated dependency version while a newer version exists. Figure~\ref{fig:pinned-dependency} shows an example of a stale pin in our previous example from the Maven library $\library{\texttt{gemini}}{\texttt{3.3.0}}$ to an outdated version of the \texttt{jackson-databind} library. When $\library{\texttt{gemini}}{\texttt{3.3.0}}$ was published, it contained a dependency to $\library{\texttt{jackson-databind}}{\version{\texttt{2.10.0}}{}}$ even though the later version $\library{\texttt{jackson-databind}}{\version{\texttt{2.11.0}}{}}$ was available. Although there was as an option to perform a semver-compatible upgrade, the consumer still kept the outdated version of the dependency. Had the library depended on $\library{\texttt{jackson-databind}}{\version{\texttt{2.11.0}}{}}$, the pin would be \textit{fresh}.

We formally define a \textit{stale pin} as follows: given a dependency graph $\mathbb{G}$, a stale pin is the tuple of three libraries \pin{$\library{C}{\version{V}{}}$}{$\library{D}{\version{V}{\alpha}}$}{$\library{D}{\version{V}{\beta}}$} $\in \mathbb{L}\times\mathbb{L}\times\mathbb{L}$ for which the following conditions hold:
\begin{enumerate}
    \item $\library{D}{\version{V}{\alpha}} \in \textit{allDeps}(\library{C}{\version{V}{}})$
    \item $\textit{publish}(\version{V}{\alpha}) < \textit{publish}(\version{V}{\beta}) < \textit{publish}(\version{V}{})$
    \item $(\textit{major}(\version{V}{\beta}) = \textit{major}(\version{V}{\alpha})) \land (\version{V}{\beta} > \version{V}{\alpha})$
\end{enumerate}

The first condition specifies that a $\library{D}{\version{V}{\alpha}}$ is a dependency of consumer $\library{C}{\version{V}{}}$. Next, the time at which each of these libraries was published is compared: if the newer dependency version $\version{V}{\beta}$ was published before the consumer version $\version{V}{}$, then consumer $\library{C}{\version{V}{}}$ contains a stale pin to dependency $\library{D}{\version{V}{\alpha}}$, as it chose to use an outdated dependency version rather than performing the upgrade to $\version{V}{\beta}$. The final condition incorporates semantic versioning guidelines and checks that the upgrade from $\version{V}{\alpha}$ to $\version{V}{\beta}$ is a semver-compatible upgrade by ensuring major version equality and using the semantic versioning ordering. This filters out any major version upgrades due to their potential of introducing backwards incompatible changes.

\change{
We then define a \textit{fresh pin} as follows: a fresh pin is the tuple of two libraries $\langle \library{C}{\version{V}{}},\library{D}{\version{V}{\gamma}} \rangle$ for which the following conditions hold:
\begin{enumerate}
    \item $\library{D}{\version{V}{\gamma}} \in \textit{allDeps}(\library{C}{\version{V}{}})$
    \item $\textit{publish}(\version{V}{\gamma}) < \textit{publish}(\version{V}{})$
    \item $\nexists \: \library{D}{\version{V}{\delta}} \in \mathbb{L}$ such that:
    \begin{enumerate}
        \item $(\textit{major}(\version{V}{\delta}) = \textit{major}(\version{V}{\gamma})) \land (\version{V}{\delta} > \version{V}{\gamma})$
        \item $\textit{publish}(\version{V}{\gamma}) < \textit{publish}(\version{V}{\delta}) < \textit{publish}(\version{V}{})$
    \end{enumerate}
\end{enumerate}
}
We can further classify a pin as either \textit{direct} or \textit{indirect} depending on the nature of the dependency between $\library{C}{\version{V}{}}$ and $\library{D}{\version{V}{\alpha}}$. \pin{$\library{C}{\version{V}{}}$}{$\library{D}{\version{V}{\alpha}}$}{$\library{D}{\version{V}{\beta}}$} is a direct if $\library{D}{\version{V}{\alpha}} \in \textit{directDeps}(\library{C}{\version{V}{}})$ and indirect if $\library{D}{\version{V}{\alpha}} \in \textit{indirectDeps}(\library{C}{\version{V}{}})$. To freshen a direct pin, a consumer would simply need to update the version of the dependency to the newer version in the project configuration file; this would change the pin from stale to fresh. Freshening indirect pins, on the other hand, requires the consumer to explicitly override the indirect dependency relation to $\library{D}{\version{V}{\alpha}}$ by introducing a new direct dependency relation to $\library{D}{\version{V}{\beta}}$. 

Freshening a pinned dependency would involve performing the upgrade from $\version{V}{\alpha}$ (pinned version) to $\version{V}{\beta}$ (upgrade version) and is not necessarily a straightforward decision. Consumers may be apprehensive of incorporating changes that break their project or even introduce new security vulnerabilities. However, keeping the dependencies pinned has a risk of missing out on crucial patches for vulnerabilities that exist in the pinned version, usually fixed in minor and patch version upgrades. Without a way of characterizing the impact of these upgrades beyond semantic versioning guidelines, developers must make a difficult decision when deciding to perform these dependency upgrades.

\section{Pinning in Maven}
\label{sec:empirical}
Using the Open Source Insights dataset published by Google~\cite{DepsDev}, we conducted an analysis on a snapshot of the Maven ecosystem to measure the frequency and impact of pinning. We take a snapshot of the entire Maven dependency network on May 22, 2023 that includes dependencies and consumers (both direct and indirect) of $\sim$567,000 Maven libraries. This snapshot contains 235,959,564 dependency edges, of which 45,997,607 (19.5\%) are direct dependencies. Ignoring different versions, there are a total of 188,927 dependencies and 377,551 consumers. \change{We analyze an older snapshot to provide time (roughly 1.5 years) for CVEs to be discovered and added to the vulnerability database}.

We chose to use the Open Source Insights dataset because it includes the dependency versions that result from Maven's dependency resolution process rather than the syntax declared in the POM files of the projects\footnote{We also considered Libraries.io~\cite{LibrariesIo} for our dataset, which stores the dependency version as the syntax of the version listed in the POM files, but chose Open Source Insights due to its explicit versioning resolution and more up to date dataset.}. This provides resolution for version ranges or keywords in the POM file (e.g., \texttt{2.0+} or \texttt{LATEST}) and also solves version conflicts for duplicate indirect dependencies (i.e., diamonds in the dependency graph). By using the final resolved versions rather than declared versions, we can find \textit{explicit} instances of stale pinning that occur in the ecosystem. \delete{To our knowledge, Open Source Insights is the most up-to-date dataset for Maven at the time of writing}.


\begin{figure}[t]
    \centering
    \includegraphics[width=0.9\columnwidth]{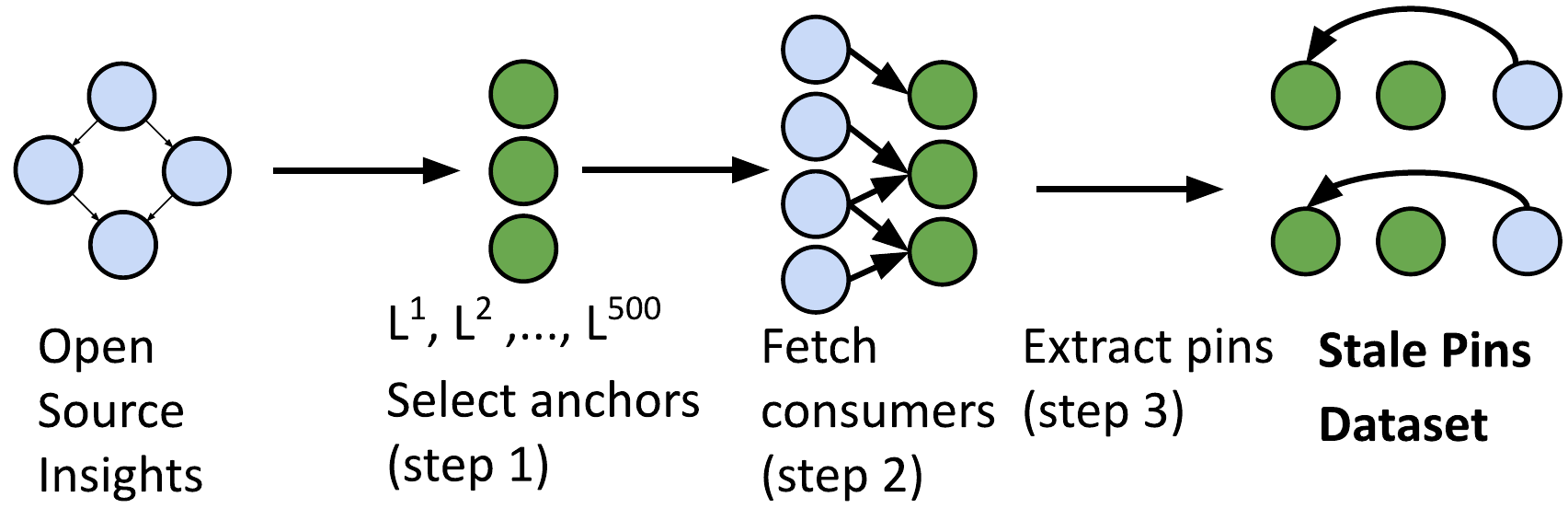}
    \caption{Construction of stale pin dataset $\mathcal{D}$. A set of anchors of selected from the Open Source Insights dataset based on popularity (number of consumers). $\mathcal{D}$ is constructed by extracting pins from the consumer of the anchors to the anchors themselves.}
    \label{fig:datasets}
\end{figure}

\subsection{RQ1: Frequency of Stale Pinning}
\label{sec:rq1}
In RQ1, we focus on how common the practice of dependency pinning is in the Maven ecosystem. Since the entire Maven ecosystem is too large to analyze in its entirety, we target our analysis to a sample of the Maven ecosystem relating to the top 500 most popular libraries (as defined by the number of consumers) due to their overall impact on the network. In particular, we analyze stale pins of consumers to this set of the most popular libraries.

We construct a dataset of stale pins (as defined in Section~\ref{sec:background:pin}) using the process shown in Figure~\ref{fig:datasets}. The dataset uses the top 500 most popular libraries (referred to as \textit{anchors}) as a starting point to find pins across the network. The anchors are created by selecting the library names (e.g., $L^1, L^2, \dots, L^{500}$) with the highest number of consumers across all versions, as seen in Step 1 of Figure~\ref{fig:datasets}.

We can walk through an example of extracting a stale pin from a consumer to an anchor. We refer back to Figure~\ref{fig:pinned-dependency} with the dependency from $\library{\texttt{gemini}}{\texttt{3.3.0}}$ to $\library{\texttt{jackson-databind}}{\texttt{2.10.0}}$. As \texttt{jackson-databind} is one of our anchors, we would like to extract stale pins from consumers to its outdated versions. We begin by querying Open Source Insights to find all the consumers of \texttt{jackson-databind}, across all versions of the library. One such consumer is \texttt{gemini}---although there are many versions of this library, we select latest minor version (\texttt{3.3.0}) to find an up-to-date version. Since there are multiple versions of \texttt{jackson-databind} higher than version \texttt{2.10.0} published earlier than $\library{\texttt{gemini}}{\texttt{3.3.0}}$, we select the highest one (\texttt{2.11.0}) and add the stale pin \pin{\library{\texttt{gemini}}{\texttt{3.3.0}}}{\library{\texttt{jackson-databind}}{\texttt{2.10.0}}}{\library{\texttt{jackson-databind}}{\texttt{2.11.0}}} to dataset $\mathcal{D}$.

The process of creating the entire dataset is formally described as follows: we first query the Open Source Insights network to find all consumers (step 2 in Figure~\ref{fig:datasets}) of the anchors libraries across all versions of each anchor, i.e. 
$$\textit{anchorConsumers} = \bigcup_{\substack{\library{L^i}{\version{V}{j}} \in \: \mathbb{L}\: \land \\
L^i \in \textit{anchors}}}  \textit{allConsumers}(\library{L^i}{\version{V}{j}})\}$$
We then query Open Source Insights to select the latest minor version of each consumer in \textit{anchorConsumers}.
For each consumer $\library{C}{V}$, we find all of its dependencies to the anchor libraries and check whether any of them are are a stale pin. Given a dependency to an anchor $\library{L^i}{\version{V}{\alpha}}$, we select the highest version $\version{V}{\beta}$ that was published before $\library{C}{V}$ and add the corresponding stale pin to $\mathcal{D}$ (step 3 of Figure~\ref{fig:datasets}).

\begin{figure}[t]
\centering
\begin{subfigure}[t]{0.23\textwidth}
\centering
\includegraphics[width=\linewidth]{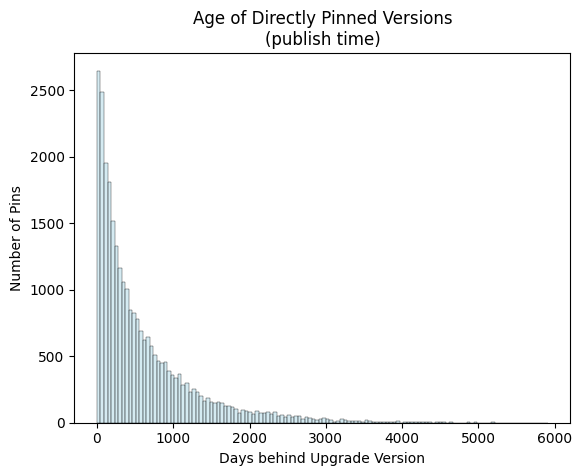}
\end{subfigure}
\begin{subfigure}[t]{0.23\textwidth}
\centering
\includegraphics[width=\linewidth]{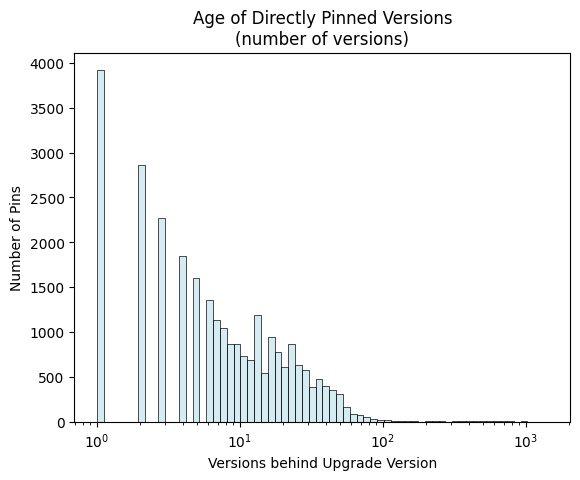}
\end{subfigure}
\begin{subfigure}[t]{0.23\textwidth}
\centering
\includegraphics[width=\linewidth]{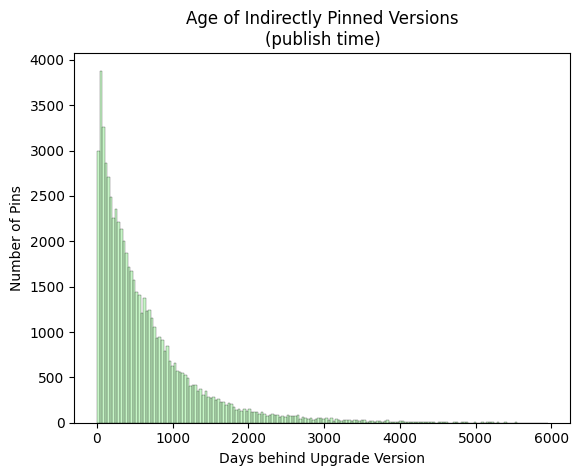}
\end{subfigure}
\begin{subfigure}[t]{0.23\textwidth}
\centering
\includegraphics[width=\linewidth]{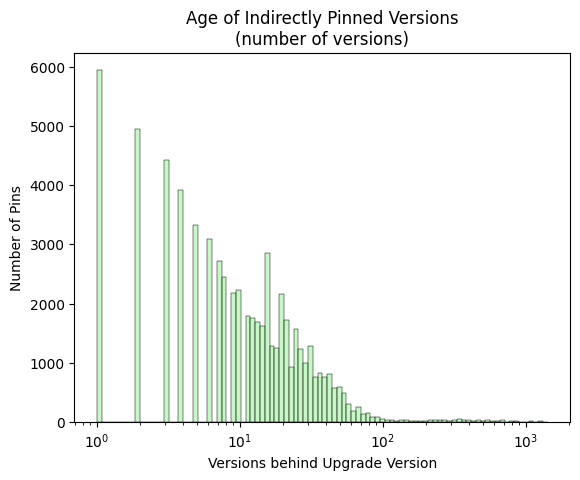}
\end{subfigure}
\caption{Histograms showing the age of direct and indirect stale pinned dependencies in $\mathcal{D}$. Direct stale pins are down in dark blue and indirect stale pins are show in light green. X-axis displays age and Y-axis displays the number of stale pins. Log scale for X-axis is used for version plots.}
\label{fig:pinned-hist-2}
\end{figure}

Table \ref{tab:dataset-2-pinning-stats} shows the statistics of the number of consumers, dependencies, and upgrades in $\mathcal{D}$. Interestingly, we find that \textit{more than 60\%} the direct consumers of the anchors contain at least 1 direct stale pin, and \textit{over 80\%} of the indirect consumers contain at least one indirect stale pin. Furthermore, we can see from Figure~\ref{fig:pinned-hist-2} that the dependency versions are outdated by a median of 370 days (7 versions) and 427 days (9 versions) for direct and indirect stale pins respectively. We see that pinning to the top 500 libraries is extremely common and features fairly outdated pinned versions! Note that there are a significantly smaller number of potential \emph{upgrades} in $\mathcal{D}$ (as defined in Section~\ref{sec:background:ecosystems}) than there are pinning consumers, suggesting that many consumers share the same stale pins to the anchors.

\mybox{\faArrowCircleRight\xspace\textbf{Finding \#1}: Stale pinning to the most popular libraries in Maven is a very common practice, with over 60\% of consumers containing at least one direct stale pin, and 80\% containing at least one indirect stale pin. The pinned versions of these libraries are fairly outdated, about 7 versions behind the upgrade version for direct pins and 9 versions for indirect pins.}


\begin{table}[t]
\centering
\caption{Pinning statistics for consumers of the top 500 most popular libraries, categorized as either direct or indirect. Out of these consumers, 148,811 (61\%) contain at least one direct stale pin and 184,281 (83\%) contain at least one indirect stale pin. We see that many consumers share the same stale pins, as there are only 46,365 potential semver-compatible upgrades in the set of direct stale pins and 76,317 potential upgrades in the set of indirect stale pins.}
\small
\begin{tabular}{|l|l|l|l|l|l|}
\hline
& \textbf{Consumers} & \makecell{\textbf{Consumers}\\\textbf{with $\ge$ 1}\\\textbf{stale pin}} & \textbf{Deps.} & \textbf{Upgrades} \\ \hline
Direct   & 244,819          &  148,811 & 717,705    &  46,365                 \\ \hline
Indirect  & 221,744          &  184,281  & 2,778,165  &  76,317                 \\ \hline
\end{tabular}
\label{tab:dataset-2-pinning-stats}
\end{table}

\subsection{RQ2: Security Impact of Stale Pinning}
Older versions of libraries frequently contain known vulnerabilities that are patched in newer minor and patch releases. These security issues are tracked and disclosed publicly using Common Vulnerabilities and Exposures (CVEs) and other reporting mechanisms. The public Open Source Vulnerabilities (OSV)~\cite{OpenSourceVulnerabilities} database maintained by Google is a central database for CVEs and is used as a data source for the Open Source Insights dataset, which stored metadata about each vulnerability as an \textit{advisory}. Each security advisory includes information about the packages and specific versions affected by the vulnerability.

From RQ1, we see that a very large percentage of consumers depend on an outdated version of the most popular libraries in the Maven ecosystem. While this provides a view of how frequent dependency pinning occurs in the Maven ecosystem, we are interested in measuring the security impact of these pins: specifically, are developers avoiding introducing new security vulnerabilities into their dependencies by pinning, or they missing out on important security patches? Tools such as \textit{dependabot} utilize these vulnerability databases to notify developers of vulnerable dependencies in the latest development snapshots; however, this data has not been used to identify the historical security impact of pinned dependencies in Maven.       

\begin{figure}[t]
\centering
\includegraphics[width=0.75\columnwidth]{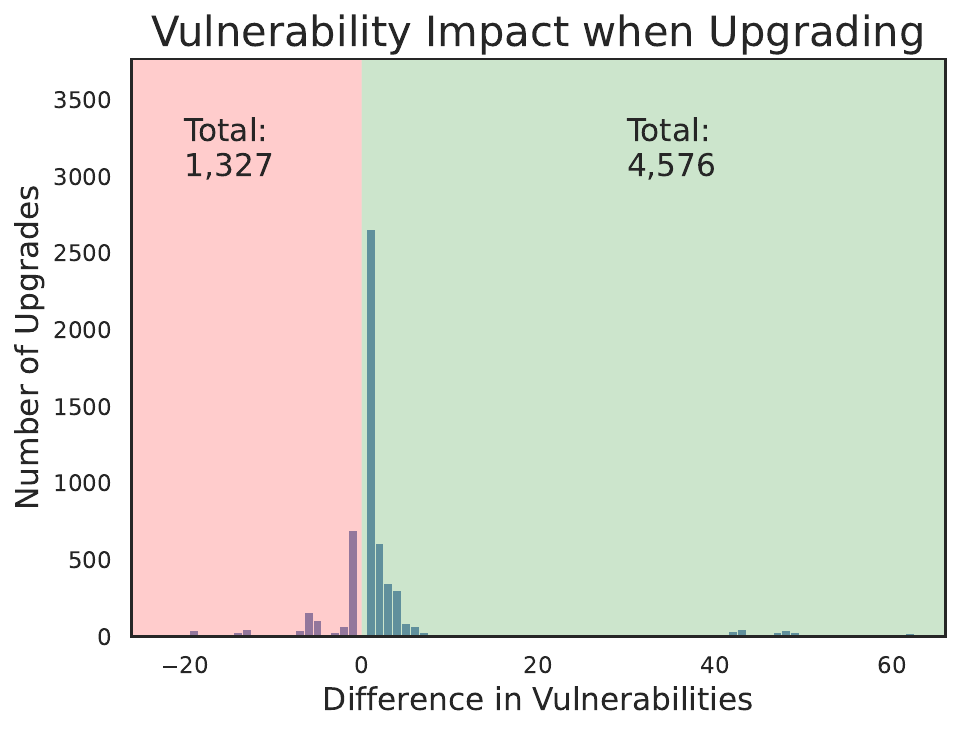}
\caption{Histogram visualizing the security advisory impact of upgrading directly pinned dependencies in $\mathcal{D}$. X-axis values in green include upgrades that reduce the number of vulnerabilities, whereas values in red increase the number of vulnerabilities (zero excluded for sake of visualization). In total, there are 4,576 upgrades that decrease vulnerabilities and 1,327 upgrades that increase vulnerabilities. Across all upgrades, the number of vulnerabilities reduce by 20,825.}
\label{fig:upgrade-advisory-hist}
\end{figure}

To perform this analysis, we compare the number of security vulnerabilities affecting the pinned version and upgrade version of the direct stale pins in dataset $\mathcal{D}$. Of the 46,365 potential upgrades (Table~\ref{tab:dataset-2-pinning-stats}), we find that 40,462 (87.2\%) result in no difference in vulnerabilities, 4,576 (9.9\%) upgrades would have reduced the number of security vulnerabilities, and 1,327 (2.9\%) would have introduced new ones. Thus, performing a semver-compatible upgrade of a pinned dependency in $\mathcal{D}$ is \textit{3.45}$\times$ as likely to fix vulnerable dependencies than introduce new ones. Figure~\ref{fig:upgrade-advisory-hist} displays the histogram of the differences in vulnerabilities between the versions, excluding the upgrades having no security impact for the sake of visualization. The majority of upgrades reduce the number of security vulnerabilities by 1, but certain upgrades can fix up to as many as 66 vulnerabilities! Across all of these upgrades, the number of vulnerabilities would be reduced by 20,825. 



\mybox{\faArrowCircleRight\xspace\textbf{Finding \#2}: 10\% of semver-compatible upgrades on a pinned version of a popular library reduced security vulnerabilities, and only 3\% introduced new ones. Although the majority of upgrades have historically featured no change in vulnerabilities, upgrading is 3.45$\times$ as likely to reduce security vulnerabilities than introduce new ones.}

\section{An Approach Towards Freshening Pins}
\label{sec:approach}

In answering RQ1 and RQ2, we have identified that stale dependency pinning to the most popular libraries in Maven is fairly common and has high security risks. However, developers of these libraries may be cautious to perform these upgrades. To upgrade a stale pin to a fresh pin, a consumer needs to be confident that the changes in the dependency upgrade are safe to introduce. One method would be to execute their own test suites against the new version of the dependency. However, even if the tests pass, they may not be comprehensive enough to thoroughly test behaviors of the new dependency version. We address this concern by developing a prototype called \emph{\tool{}} that calculates a confidence score of a given upgrade by executing \textit{crowdsourced} test suites of other consumers of the pinned dependency and measuring their outcomes. Our insight is that consumer test suites can exercise a more thorough set of behaviors of the dependency; if multiple consumers' tests pass on both the stale pin and fresh pin, \tool{} can provide additional signal for a developer to more confidently upgrade their dependency.
 
\begin{figure}[t]
    \centering
    \includegraphics[width=0.9\columnwidth]{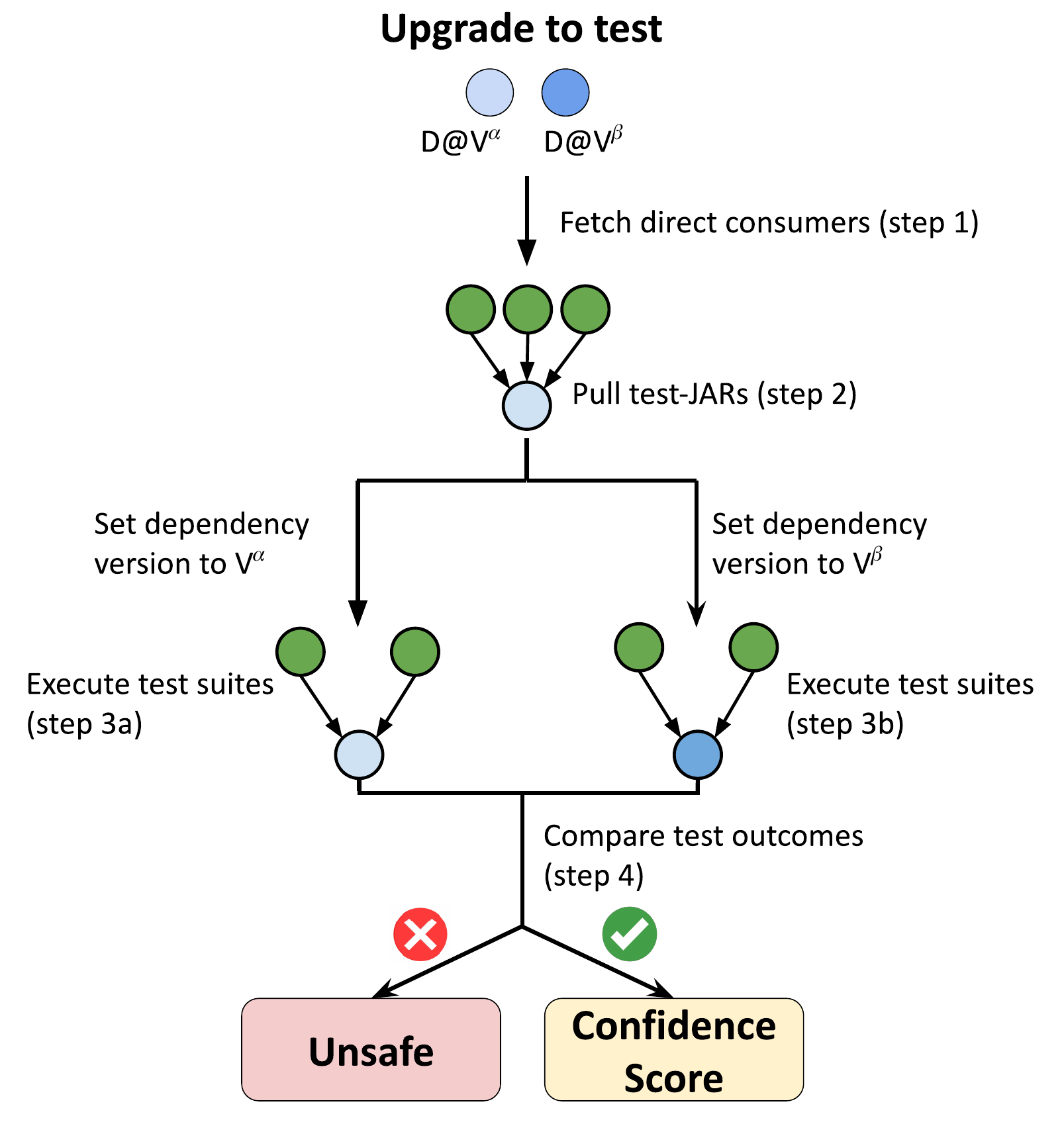}
    \caption{Overview of \tool{}. 
    The direct consumers of $\library{D}{\version{V}{\alpha}}$ are fetched from Open Source Insights and their test suites are executed using test-JARs. Then, the version is set to $\version{V}{\beta}$ and the consumer test suites are executed on this version. Finally, the test outcomes are compared---either the upgrade is safe and \tool{} returns a confidence equal to the number of test suites executed, or the upgrade is unsafe, returning zero confidence.}
    \label{fig:unpin-procedure}
\end{figure}

\tool{} takes an upgrade $(\library{D}{\version{V}{\alpha}}, \library{D}{\version{V}{\beta}})$ as input and crowdsources test suites to calculate a confidence score for the upgrade. Our approach follows the procedure outlined in Figure~\ref{fig:unpin-procedure}:
\begin{enumerate}
\item Query Open Source Insights to find $\textit{directConsumers}(\library{D}{\version{V}{\alpha}})$.
\item Pull the consumer test-JARs from the Maven Central Repository for each $\library{C}{V} \in \textit{directConsumers}$. Note that not all consumers have published test-JARs; thus, we construct a set $\textit{testableConsumers} = \{\library{C}{V} \in \textit{directConsumers}(\library{D}{\version{V}{\alpha}}) \:|\: \textit{testJarExists}(\library{C}{V})\}$
\item For each consumer $\library{C}{V} \in \textit{testableConsumers}$, execute the tests when using $\library{D}{\version{V}{\alpha}}$ (3a) and $\library{D}{\version{V}{\beta}}$ (3b) as dependencies (see Section~\ref{sec:tests}). 
\item Compare the test outcomes for each version and calculate a confidence for the upgrade \upgrade{\library{D}{\version{V}{\alpha}}}{\library{D}{\version{V}{\beta}}} (see Section~\ref{sec:confidence}).
\end{enumerate}

Steps (1) and (2) query the Open Source Insights dataset and the Maven Central Repository respectively to fetch test-JARs of the direct consumers of the pinned version. In the following sections, we go into detail to describe Steps (3) and (4). 

\subsection{Executing Consumer Test Suites}

\label{sec:tests}
\label{sec:approach:testsuites}


One option to execute a consumer test suite is to download and build the source code of the repository and invoke the tests by running \texttt{mvn test}. Unfortunately, the source code for these consumers may not be publicly available. Additionally, resolving the specific version $V$ in the repository can be a nontrivial task, as version naming conventions may differ between the source code and the Maven package.    

The strategy we chose was to leverage the Maven Central Repository for \textit{test-JARs} of the consumer, which contains compiled classes of the test files. Test-JARs are unique to the Maven ecosystem and provide a streamlined approach of fetching and executing project test suites. While test-JARs are optional to upload to the Maven Central Repository and do not exist for certain consumers, this approach still provides a straightforward method of crowdsourcing test suites. Among the consumers in $\mathcal{D}$ with direct stale pins, we found that about 12\% of projects had uploaded test-JARs to the Maven Central Repository; while we would have liked this percentage to be higher, this is still a significant number of tests available for \tool{} to use to test upgrades.

To walk through this process, we refer to our original example of a pinned dependency from \texttt{gemini@3.3.0} to \texttt{jackson-databind@2.10.0}. The consumer \texttt{gemini@3.3.0} would use \tool{} to test the upgrade of \texttt{jackson-databind} from \texttt{2.10.0} to \texttt{2.11.0}. \tool{} first finds all consumers of the pinned dependency \texttt{jackson-databind@2.10.0} and pulls all consumer test-JARs that are available on the Maven Central Repository. In the case that there are multiple consumers with the same library name, we select the highest version.

For each of the consumers, \tool{} first executes each of the test suites against the pinned dependency version of \texttt{jackson-databind} (\texttt{2.10.0}). Some tests may produce non-deterministic outcomes due to \textit{flakiness}~\cite{luo2014empirical, eck2019understanding}. \tool{} executes each test with $r = 5$ repetitions to account for this flakiness. Since the tests are executed directly from the test-JARs, it also is possible that tests may have errors or fail due to missing resources. We save the test outcomes produced by Maven of each of the consumer tests to a database.

Next, \tool{} upgrades the dependency version of \texttt{jackson-databind} to \texttt{2.11.0} for each of the consumer test suites. Once again, the execution of the test suites are repeated five times, and the test outcomes are saved.

\begin{figure}[t]
\centering
\begin{subfigure}[t]{0.49\columnwidth}
\centering
\includegraphics[width=\linewidth]{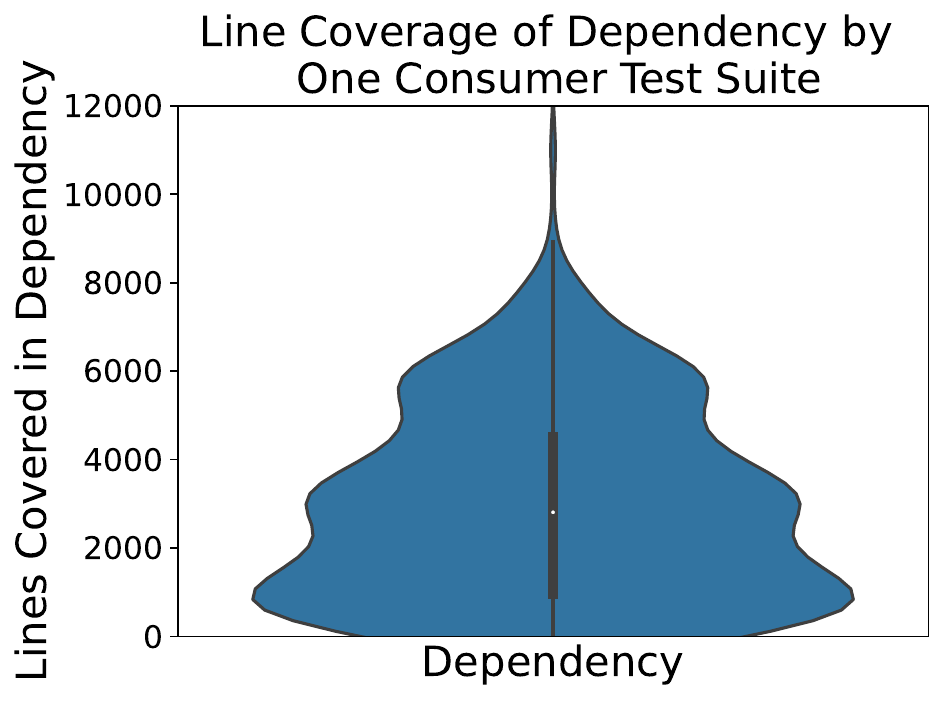}
\end{subfigure}
\begin{subfigure}[t]{0.49\columnwidth}
\centering
\includegraphics[width=\linewidth]{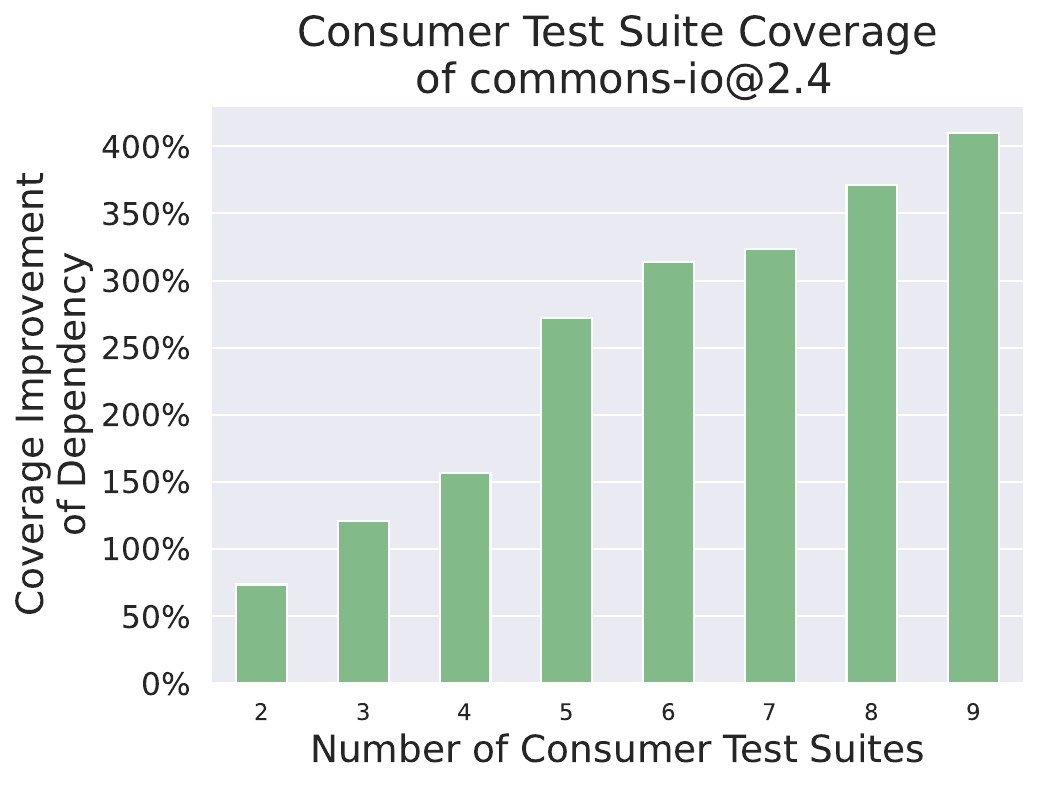}
\end{subfigure}
\caption{The left plot displays a violin plot displaying distribution of average number of lines covered for a dependency from a \textit{single} consumer test suite. The right plot displays the line coverage \textit{improvement} of \library{\texttt{commons-io}}{\texttt{2.4}} from additional consumer test suites, measured using random sampling.}
\label{fig:raw-cov-and-cov-improvement-commons-io}
\vspace{-1.5em}
\end{figure}



\subsection{RQ3: Coverage Improvement of Crowdsourced Consumer Test Suites}

A natural question is whether using additional consumer test suites has any advantages in terms of exercising code, such as improved coverage, of the dependency. To characterize the coverage benefit of crowdsourced consumer test suites, we use the Jacoco library~\cite{jacoco} to collect the coverage of the dependency classes only. Figure~\ref{fig:raw-cov-and-cov-improvement-commons-io} (left) shows the coverage distribution of dependencies by a single consumer test suite (on average). We observe that consumer test suites cover a substantial amount of dependency code, with some even covering up to 10,000 lines. Figure~\ref{fig:raw-cov-and-cov-improvement-commons-io} (right) shows the coverage \textit{improvement } from executing additional consumer test suites on one pinned dependency \texttt{commons-io@2.4}---\tool{} finds nine consumers of this dependency whose test JARs could be executed. From the figure, we see that the union line coverage of these nine test suites provided over a 400\% increase in coverage of \texttt{commons-io@2.4} than if we only executed a single consumer's test suite (on average). To understand how coverage increases with the number of crowdsourced test suites, we calculate the union of the dependency coverage for each value $n$ below 9 by randomly sampling a subset of $n$ consumer test suites without replacement (up to 50 times) and calculating the average.

\begin{figure}[t]
\centering
\includegraphics[width=0.7\columnwidth]{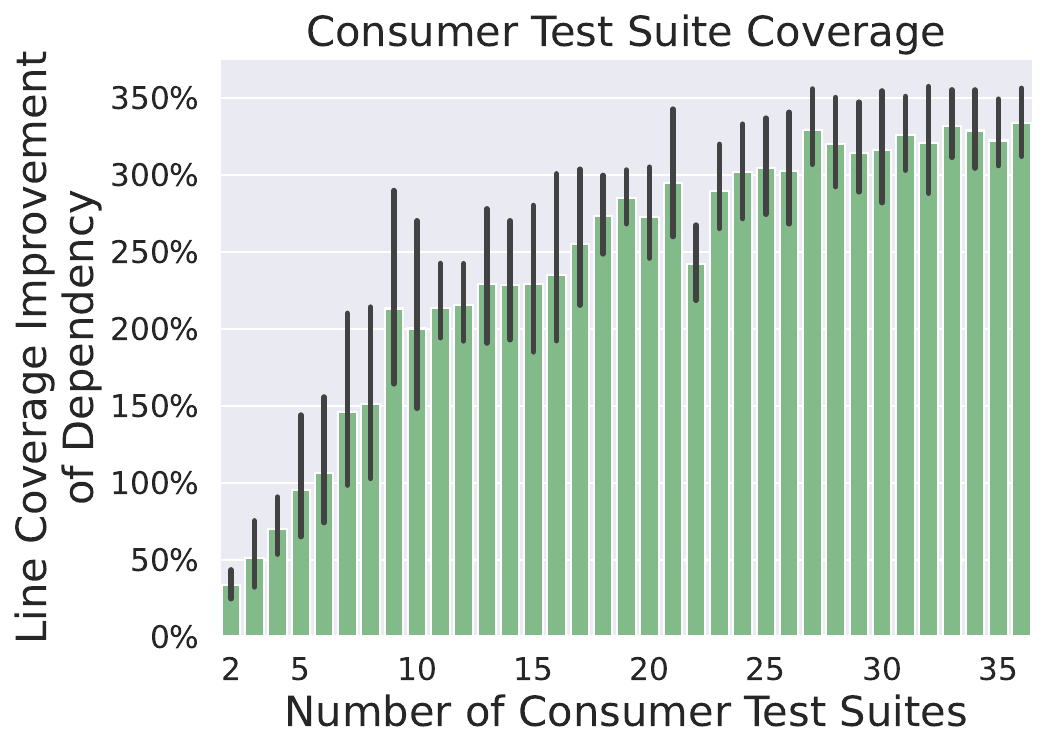}
\caption{Average coverage improvement achieved by \tool{} over an average consumer test suite (higher is better). X-axis values include the number of crowdsourced consumer test suites, and Y-values show the geometric mean line-coverage improvement across all libraries. As low as four additional crowdsourced test suites can achieve almost 100\% more line coverage than a single one. }
\label{fig:cov-improvement}
\end{figure}

Generalizing this methodology, Figure~\ref{fig:cov-improvement} 
shows the average coverage improvement, across all popular libraries, with respect to the number of consumer test suites. With just a single additional consumer test suite, we can achieve an average of 40\% additional coverage of the dependency; with four additional test suites, this number rises to almost 100\%! Overall, we find that the crowdsourced test suites from \tool{} are able to gain a significant coverage boost in the pinned dependency over a single consumer, thus providing more confidence in an upgrade.

\subsection{Computing Confidence Score}
\label{sec:confidence}

We next explain how \tool{} uses the outcomes from the consumer test suites to test an upgrade. Based on the results of the crowdsourced test suites, \tool{} calculates a \textit{confidence} score for each upgrade. We walk through our example of upgrading \texttt{jackson-databind} from version \texttt{2.10.0} to \texttt{2.11.0}. There are seven available consumer test-JARs, which \tool{} executes on the pinned version \texttt{2.10.0} and upgrade version \texttt{2.11.0}. Tests that are flaky or fail in the pinned version are filtered out, and all remaining test outcomes are compared between versions. Each of the seven consumers \textit{vote} on whether the upgrade is safe or unsafe. If all tests in the consumer test suite pass on both dependency versions, then that consumer votes \emph{safe}; otherwise, there exists a test that passes in the pinned version but fails in the upgrade version, indicating the presence of a breaking change. Since all seven consumers vote safe, the confidence returned by \tool{} is seven, indicating the presence of seven passing consumer test suites. Had there been a consumer that voted unsafe, then the confidence returned by \tool{} would be zero.

More formally, we determine confidence as follows. We define \textit{outcome} as a function that takes in a test method $t$, a consumer $\library{C}{\version{V}{}}$, and a dependency $\library{D}{\version{V}{}}$. From Section~\ref{sec:tests}, each test has been executed with $r$ repetitions.

\[ \textit{outcome}(t, \library{C}{\version{V}{}}, \library{D}{\version{V}{}}) = 
\begin{cases}
\textit{pass} & \text{if $r$ repetitions pass } \\ 

\textit{fail} & \text{if $r$ repetitions fail } \\
\textit{flaky} & \text{otherwise}  
\end{cases}
\]


Each consumer provides a \textit{vote} for whether the upgrade is safe or unsafe depending on the results of its test suite. If all passing tests with dependency version $\version{V}{\alpha}$ also pass when the dependency version is upgraded to $\version{V}{\beta}$, then the consumer vote is \textit{safe}. If there is a test that consistently passes with $\version{V}{\alpha}$ but always fails with $\version{V}{\beta}$, then the consumer vote is \textit{unsafe}---this condition indicates that the upgrade has broken some functionality. In all other cases (e.g., all tests were flaky or failed in $\version{V}{\alpha}$), the consumer vote is ignored.  

\begin{align*}
&\textit{vote}(\library{C}{\version{V}{}}, \library{D}{\version{V}{\alpha}}, \library{D}{\version{V}{\beta}}) = \\ &\begin{cases}
\textit{safe} & \text{if}\: \forall \: t \in \textit{consumerTests}(\library{C}{\version{V}{}}) : \\ & \qquad \textit{outcome}(t, \library{C}{\version{V}{}}, \library{D}{\version{V}{\alpha}}) = \textit{pass} \implies \\ & \qquad \textit{outcome}(t, \library{C}{\version{V}{}}, \library{D}{\version{V}{\beta}}) = \textit{pass} \\ 
\textit{unsafe} & \exists \: t \in \textit{consumerTests}(\library{C}{\version{V}{}}) : \\ & \qquad \textit{outcome}(t, \library{C}{\version{V}{}}, \library{D}{\version{V}{\alpha}}) = \textit{pass} \: \land \\ & \qquad \textit{outcome}(t, \library{C}{\version{V}{}}, \library{D}{\version{V}{\beta}}) = \textit{fail} \\
\textit{ignore} & \text{otherwise}
\end{cases}
\end{align*}
where $\textit{consumerTests}(\library{C}{\version{V}{}})$ returns the set of all test methods in the test-JAR for $\library{C}{\version{V}{}}$. 

Finally, \tool{} accumulates all votes of the consumers to calculate a \textit{confidence} score for the upgrade. If any consumers vote that the upgrade is unsafe, then the confidence is 0, since the upgrade appears to be a breaking change. Otherwise, the confidence is equal to the number of consumers that voted \textit{safe}---higher is better. 



The confidence score calculated by \tool{} reports the number of consumers that had consistent test results between dependency versions. This score does not provide any guarantees about the safety of the upgrade---it is possible that the executed consumer test suites did not catch a breaking change. However, even without guarantees, this score is an improvement over the standard technique of just using one's own tests. We also note that \tool{} provides a \textit{conservative} estimate of safety by reporting a score of 0 if any of the consumer test suites fail. However, the interpretation of the score depends on the preferences of the developer looking to perform the upgrade. The confidence scores reported \tool{} will also increase with more testable consumers and more available test-JARs.


\subsection{RQ4: Providing Additional Signal for Freshening Pins}

A key question is how much additional signal \tool{} can provide to consumers to upgrade one or more of their dependencies. We answer this RQ by running \tool{} on the upgrades of direct stale pins in $\mathcal{D}$ that fix security vulnerabilities. Table~\ref{tab:unpin-confidence} reports the distribution of upgrades that had a positive and zero confidence returned by \tool{}. About 29\% of all upgrades were able to be tested with at least 1 test-JAR crowdsourced from the Maven Central Repository. Out of these tested upgrades, \tool{} reported a positive confidence score for 850 (65\% of tested upgrades, 19\% of all upgrades). In total, 9,194 (41\%) consumers contained these stale pins and would have a positive signal from \tool{} to perform these vulnerability-fixing upgrades. 

\begin{figure}[t]
\centering
\includegraphics[width=0.8\columnwidth]{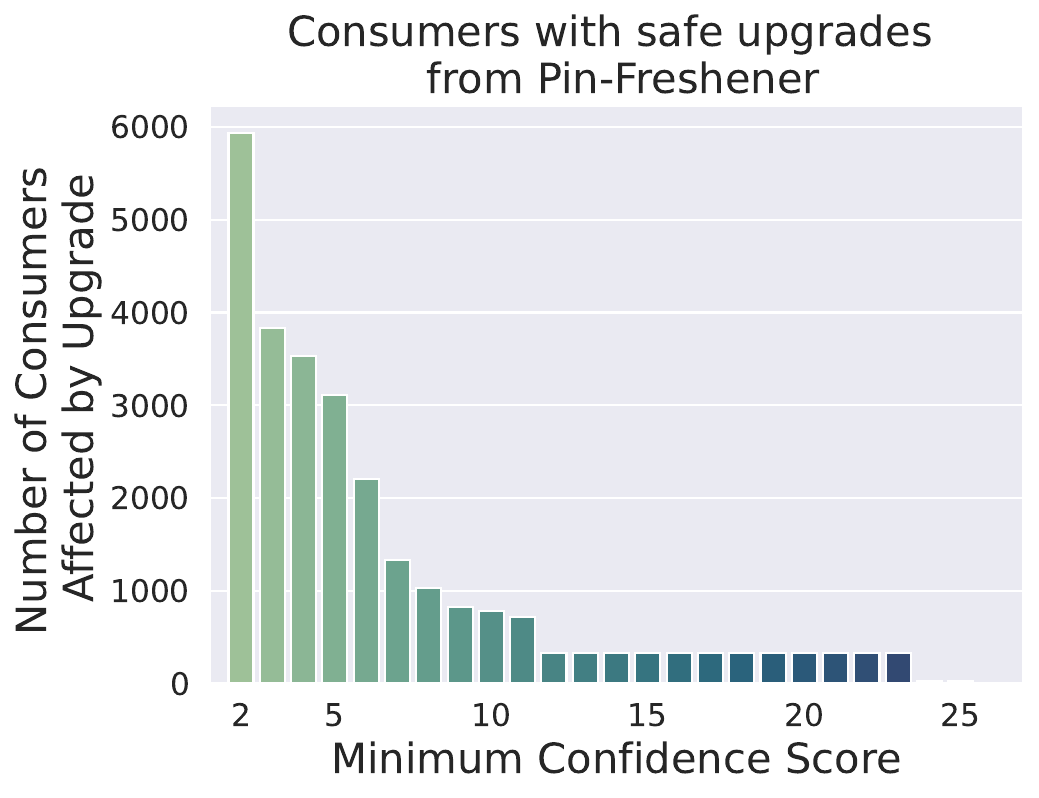}
\caption{Number of consumers with positive score upgrades with respect to confidence score returned by \tool{}. X-axis displays the minimum number of passing test suites (1 is excluded for visualization), and Y-values are the number of consumers that would be able to upgrade given the confidence value. Over 3,000 consumers have upgrades with at least 5 safe consumers, and almost 6,000 with at least 2.}
\label{fig:unpin-safe}
\end{figure}

For these 9,914 consumers with positive confidence scores from \tool{}, what is the distribution of these scores? Figure~\ref{fig:unpin-safe} visualizes these consumers against values of the confidence score. The X-axis value of 1 is excluded for the sake of visualization and because we believe a minimum score of 1 is too low to provide enough signal for an upgrade. Overall, we find that over 3,000 (14\%) of consumers would be provided a confidence score of at least 5 using \tool{}. This number of consumers increases to almost \textit{6,000} with a confidence score of at least 2. This is a significant number of consumers that would be provided additional signal to upgrade their pinned dependencies with these consumers validating the upgrade. We believe this number can be increased even further with more Maven libraries adopting the practice of publishing their test-JARs. 

\begin{table}[t]
\caption{\tool{} confidence on upgrades of direct pins $\mathcal{D}$ that reduce security vulnerabilities and the number of consumers affected. Out of the 4,576 upgrades, \tool{} was able to crowdsource at least one test-JAR for 29\% (upgrades with zero and positive confidence). \tool{} returns a positive confidence for 9,194 (41\%) of all consumers that could have performed these upgrades.} 
\small
\begin{tabular}{|l|l|l|}
\hline
\makecell{\textbf{Confidence score}\\\textbf{from \tool{}}} & \textbf{Consumers} & \textbf{Upgrades} \\ \hline
Positive (potentially safe) & 9,194 (41\%)         &  850 (19\%)                         \\ \hline
Zero (potentially unsafe) & 3,134  (14\%)       &  458 (10\%) \\ \hline
Untested (no test-JARs) & 10,119 (45\%) & 3,268 (71\%) \\ \hline
Total & 22,447 (100\%) & 4,576 (100\%) \\ \hline
\end{tabular}
\label{tab:unpin-confidence}
\vspace{-1em}
\end{table}

\section{Discussion}
\label{sec:discussion}

In this section, we discuss our findings and their broader implications to practitioners and researchers.

\paragraph{Stale dependency pinning is common in Maven} From our analysis of dependency pinning in Maven, we find that stale pinning is fairly common for consumers of popular libraries. This is likely because popular libraries have more maintainers that can manage dependencies and keep them up to date. Further, the default behavior in Maven for specifying single dependency versions is to pin, whereas the default notation in other ecosystems like npm is a floating dependency. It can be challenging for consumers to stay up to date with the frequent releases of popular libraries. While our analysis focuses on \textit{explicit} instances of dependency pinning in the network, our findings are consistent with the studies showing how developers are reluctant to upgrade their dependencies~\cite{cox2015measuring, kula2015trusting, kula2018developers, dietrich2019dependency}. 

\paragraph{Fresh pinning is safer than stale pinning} Our historical analysis of pinned dependencies to popular libraries shows that upgrading pinned version would have had a large security impact across the ecosystem. Although consumers may be inclined to stick to a consistent dependency version, they are far likelier to fix critical security vulnerabilities by keeping their dependencies up to date. This aligns with previous studies demonstrating correlations between outdated dependencies and vulnerabilities~\cite{pashchenko2018vulnerable}. While we understand the complicated decision making that is involved in performing these upgrades, we hope developers are encouraged to adopt a more progressive strategy of keeping dependencies up to date.  

\paragraph{Coverage of a dependency improves with crowdsourced test suites} It is challenging for a consumer to evaluate how their project will be affected by a dependency upgrade. While their own test suites may be able to catch certain issues, we see that \textit{crowdsourcing} test suites from other consumers can provide a substantial boost in coverage. These test suites may be exercising different parts of the dependency, and a consumer may only care about a certain functionality that they use; nevertheless, we feel each additional test suite can only help in increasing confidence for an upgrade. Prior work has shown the potential for consumer tests ~\cite{mezzetti2018type, chen2020taming, mujahid2020using, hejderup2022can} in achieving reasonable coverage and fault detection capabilities in dependencies.

\paragraph{Ecosystems should encourage developers to publicize executable test suites} Our tool \tool{} leverages the published test-JARs in the Central Maven Repository. We believe this is a great practice to improve the overall testing infrastructure in the ecosystem and hope to see it more widely adopted by other libraries. In particular, the existence of test-JARs in the Central Maven Repository allows \tool{} to streamline the automatic execution of these tests. This infrastructure is extremely valuable and hope to see it in other ecosystems beyond Maven/Java as well. Our approach of using external test suites to validate dependency changes is similar to how \textit{monorepo} environments operate in large companies~\cite{memon2017taming} in which tests from external modules are selected and run to validate code changes. \tool{} applies this idea to the open source world through the execution of consumer test suites, providing something akin to a "monorepo for the masses". 

\paragraph{Limitations}{While \tool{} is able to provide additional signal of crowdsourced test suites to measure safety of upgrades, we acknowledge that this signal does not provide any \textit{guarantee} of safety for a specific consumer. We believe that this type of signal, however, would still be useful for developers to understand how the upgrade impacts the rest of the ecosystem and would be beneficial in making a decision about whether to upgrade from a stale to fresh pin. This is similar to how test coverage is not a guarantee of the absence of bugs but is still a widely used measure of confidence. More coverage is always good; confidence increases monotonically even if it never reaches 100\%. Our score provides a similar rating for developers familiar with such a notion of test confidence.}
\section{Threats to Validity}
\paragraph{Threats to Construct Validity}
The validation performed by \tool{} on an upgrade is dependent on the consumer tests that are executed. If there is any noise or nondeterminism affecting the test outcome, then \tool{} may improperly classify certain upgrades as safe or unsafe.  This can arise from flakiness~\cite{luo2014empirical, lam2020study, parry2021survey} in tests. We aim to mitigate this threat through repeated execution of the tests five times (Section~\ref{sec:tests}) on both the pinned version and the upgrade version. \tool{} only compares tests that produce a consistent passing or failing outcome across all repetitions, which should filter out a majority of flaky tests.

\paragraph{Threats to Internal Validity}
\tool{}'s approach of crowdsourcing test suites and validating upgrades assumes that consumer test suites are a valuable source testing a dependency. Since library test suites are generally focused on testing functionality of the library and not the dependencies, it may be the case that consumer tests do not exercise much behavior of dependencies. Nevertheless, \tool{} executes as many consumer test suites as are available in the Maven Central Repository. We hope that publishing test-JARs becomes a more widely adopted practice in Maven, as this would increase the overall coverage of the dependency.

\paragraph{Threats to External Validity}
We specifically focused on the Maven ecosystem for our analysis, and we do not know if our conclusions about dependency pinning and its security implications will generalize to other ecosystems. Additionally, \tool{} depends on a central repository of crowdsourced tests that can be automatically executed; this data may not always be available in other platforms. 
\section{Related Work}
\subsection{Dependencies in Software Ecosystems}
The challenge of evolving and maintaining software in ecosystems is a well-researched topic~\cite{lehman1998implications, lehman2005role, godfrey2008past, d2008supporting, rajlich2014software,  tripathy2014software, chapin2001types, mens2005challenges, mens2008analysing, manikas2013software, cox2019surviving, zhou2024revealing}. Bavota et al.~\cite{bavota2015apache} explore the Apache ecosystem and highlight the exponential growth in the number dependencies. They also found that application developers are reluctant to upgrade their dependencies due to the risk of API breaking changes. This issue is further quantified by Kula et al. (2015)~\cite{kula2015trusting}, sampling 4.6K Github projects and finding that more than 80 percent of them have outdated Maven dependencies. Additional studies~\cite{zerouali2018empirical} validate this finding for other ecosystems such as NPM by measuring technical lag in dependencies. Dietrich et al.~\cite{dietrich2019dependency} demonstrate that 85.7\% of Maven libraries specify a fixed version in dependencies---our definition of stale pinning is more precise as it compares the resolved version to the latest dependency version available at the time of publishing. Cox et al.~\cite{cox2015measuring} conduct a mixed methods study on outdated dependencies in Maven and propose a system-level metric for dependency freshness; our analysis of stale pins confirms that outdated dependencies are frequent even in recent snapshots of the Maven ecosystem and we provide \tool as a method for encouraging developers to upgrade to fresh pins. 

Prior work~\cite{liu2022demystifying, decan2019empirical} has also measured the impact of vulnerabilities in dependencies in the NPM ecosystem. Kula et al. (2018)~\cite{kula2018developers} extend their work to study the extent to which developers upgrade their dependencies and the reasons behind their reluctancy~\cite{kula2018developers}. In a survey of developers, they find that 69\% claimed to be unaware of vulnerabilities in their dependencies. He et al.~\cite{he2025pinning} perform an empirical study on the npm ecosystem and show that dependency pinning can increase the attack surface of malicious package updates. Automated dependency management bots like \textit{Dependabot}~\cite{Dependabot} are able to address this issue by automatically notifying and creating pull requests for developers to upgrade their vulnerable dependencies. Analysis on Dependabot in practice shows that it does reduce technical lag in projects; however, its compatibility score does not reduce developer suspicion when performing upgrades~\cite{he2023automating}. Our approach can provide additional signal through the execution of consumer test suites. 

\subsection{Detection of Breaking Changes} 
Prior research has studied~\cite{bogart2016break, xavier2017historical, bogart2021and} and developed numerous techniques for the detection of breaking changes~\cite{brito2018apidiff, du2022aexpy, ochoa2022breaking} that can alert developers of unsafe upgrades. 
\paragraph{Static Analysis Based Techniques} The majority of existing literature focuses primarily on detection of API changes between library versions. Raemaekers et al.~\cite{raemaekers2014semantic} utlilize the tool \texttt{clirr} to detect API binary incompatibilities of Java code through static analysis. APIDiff is a tool developed by Brito et al.~\cite{brito2018apidiff} that focuses on syntactic changes between Java library versions that classifies a code change as breaking or non-breaking. The more recent tool Sembid~\cite{zhang2022has} locates breaking changes in Maven libraries by analyzing call chains and measuring semantic differences between versions. 
\paragraph{Dynamic Analysis Based Techniques} Mostafa et al.~\cite{mostafa2017experience} study the prevalence of \textit{behavioral} backwards incompatibilities (BBIs) in consecutive versions of Java libraries. They find that 14 of the 15 subjects featured these types of breaking changes, with the majority of them undocumented. Prior work has also shown the effectiveness of using consumer tests to detect breaking changes and BBIs~\cite{zhang2022has, mujahid2020using, mezzetti2018type}. We highlight the main differences from our work: first, we provide a novel definition of explicit dependency pins and present a thorough empirical study on pinning in the Maven network, which is unique among related work. We also use a dataset that resolves dependency versions for old libraries at the time they were built; this is contrast to prior work that uses heuristics to resolves dependencies in older releases~\cite{mujahid2020using}. We focus on the security impact of pinning dependencies and validating upgrades from pins, which is unique among related work. Finally, we use crowdsourced tests from JARs published to the Maven central repository, and thus do not rely on identifying source code repositories like prior work~\cite{mujahid2020using, chen2020taming, mezzetti2018type}. A potential future extension of our work is to use generative AI and agentic techniques to automatically build repositories and run tests~\cite{bouzenia2024you}.

\subsection{Client-Specific Testing}
Previous techniques have aimed to measure the effect of changes in dependency components on clients. Mora et al.~\cite{mora2018client} develop an automated technique to explore behaviors of the client through symbolic execution. Zhu et al.~\cite{zhu2023client} develop Compcheck, which leverages an offline knowledge base of incompatibility issues to find similar library usages and generate tests revealing client incompatibility issues. We believe these techniques can further enhance the use of \tool{}, as they can provide more detailed analysis on test execution results; however, they require heavier-weight analysis on test executions and source code. Our approach provides a pragmatic middle-ground to execute all client test-JARs for a given library version with coverage analysis as a signal for developers. Prior automated test generation techniques to generate exploits for vulnerabilities in dependencies \cite{kang2022test,chen2023exploiting} can be for additional testing in the ecosystem that could be crowdsourced.


\section{Conclusion} 
In this work, we focused on the issue of dependency pinning in the Maven ecosystem. We introduced the concepts of \textit{stale} and \textit{fresh} pins and conducted an analysis on a recent snapshot of the Maven ecosystem, identifying that a significant portion of consumers are pinned to older versions of the most popular libraries. We also show that consumers are more likely to fix existing security vulnerabilities than introduce new ones if they were to upgrade their stale pins to fresh pins. To encourage developers to upgrade dependencies, we prototype \tool{}, an approach to execute crowdsourced consumer test suites in order to measure safety of an upgrade. Pin-Freshener can provide confidence through crowdsourced test suites to over 19\% of consumers who could fix vulnerabilities by upgrading dependencies. This is an improvement of the standard approach of just relying on executing one's own tests. We argue that more libraries and package management platforms should adopt the practice of publishing executable test binaries for further development of tools that leverage information about consumer test suites.
\section{Data Availability}
We have included evaluation data in the repository at: \url{https://doi.org/10.5281/zenodo.15021893}. This data contains dependency data for each of the datasets, coverage data for consumer test suites, and test outcome data from \tool{}.

\section{Acknowledgements}
This research was funded in part by NSF grant CCF-2120955 and the Future Enterprise Security Initiative at CyLab.

\bibliographystyle{IEEEtran}
\bibliography{IEEEabrv,references}
\end{document}